\documentclass{elsart}

\usepackage{amssymb}

\begin{document}

\begin{frontmatter}



\title{Exact Solutions of Holonomic Quantum Computation}


\author[label1]{Shogo Tanimura\corauthref{cor}}
\corauth[cor]{Corresponding author.}
\ead{tanimura@mech.eng.osaka-cu.ac.jp},
\author[label2]{Daisuke Hayashi}
\ead{daisuke{\_}hayashi{\_}0102@nifty.com},
\author[label3]{Mikio Nakahara}
\ead{nakahara@math.kindai.ac.jp} 
\address[label1]{
Graduate School of Engineering, Osaka City University,
Osaka, 558-8585, Japan
}
\address[label2]{
Department of Engineering Physics and Mechanics,
Kyoto University, Kyoto, 606-8501, Japan
}
\address[label3]{
Department of Physics, Kinki University,
Higashi-Osaka, 577-8502, Japan
}

\begin{abstract}
% Text of abstract
Holonomic quantum computation is analyzed from geometrical viewpoint.
We develop an optimization scheme in which an arbitrary unitary
gate is implemented with a small circle in a complex projective space.
Exact solutions for the Hadamard, CNOT and 2-qubit discrete Fourier
transformation gates are explicitly constructed.
\end{abstract}

\end{frontmatter}

% main text
% \section{}
% \label{}
%\section{Introduction}

Recently quantum computer attracts great interests from many
disciplines.
It is strongly desired to find a scheme to implement unitary gates in a physical system. For this purpose it is natural 
to consider utilizing a quantum system described by a Hamiltonian
$ H(\lambda) $ that depends on external parameters $\{ \lambda \}$. 
In holonomic quantum computation (HQC) 
proposed by Zanardi and Rasetti \cite{first,pachos1}, in contrast,
the holonomy \cite{wz} associated
with adiabatic change of the parameters along a loop
in a control parameter manifold is employed to implement a unitary gate.
Experimental schemes to manipulate the non-Abelian holonomy
have been proposed \cite{unanyan}
and their uses to realize a unitary gate are also proposed \cite{duan,pachos2,faoro}.
For efficient achievement of holonomic computation,
it is necessary to find a loop as short as possible in the control manifold.
A numerical scheme to search the shortest loop 
is being developed in \cite{ours1,ours2} to implement an arbitrary gate.
% implemented in HQC by searching the loop numerically.
In this Letter we consider an ideal quantum system that has full 
isospectral parameters as control parameters.
In the following we make exact analysis of the shortest loops 
that generate well-known unitary gates as holonomies.
We would like to emphasize that to find an exact solution has remained unsolved \cite{mont} even for such an idealized system.

Here we take the relevant terminology from \cite{gtp,fujii1,fujii2} 
to define our model.
Our formulation can be compared with \cite{pachos3}.
Suppose the Hilbert space of a quantum system 
is an $ N $-dimensional complex space $ \mathbb{C}^N $.
By isospectral parameters, 
we mean a family of unitary transformations $ g(\lambda) \in U(N) $
since the transformed Hamiltonians
$ H ( \lambda ) = g(\lambda) H_0 \, g^\dagger (\lambda) $
has spectra independent of $\lambda$.
We assume that
the ground state of the reference 
Hamiltonian $ H_0 $ is $ k $-fold degenerate, taking a diagonal form
\begin{equation}
	H_0 = {\rm diag} 
	( \varepsilon_1, \varepsilon_2, \ldots, \varepsilon_N )
\end{equation}
with $ \varepsilon_1 = \varepsilon_2 = \cdots = \varepsilon_k < \varepsilon_{k+j} $
$ ( j=1, \ldots, N-k) $. The ground state energy may be put to zero,
without loss of generality, to get rid of the dynamical phase factor.
We concentrate our attention on the ground states of the Hamiltonians 
$ \{ H(\lambda) \} $.
% mark
Associated with the lowest energy for each $ H(\lambda) $,
there are $ k $ orthonormal state vectors 
$ \{ |v_1 (\lambda) \rangle, \ldots, |v_k (\lambda) \rangle \} $.
The set of $ k $ orthonormal state vectors is called a $ k $-frame
and the set of all the $ k $-frames,
\begin{equation}
	S_{N, k} (\mathbb{C}) 
	= \{V \in M(N, k; \mathbb{C}) \, | \, V^{\dagger} V = I_k\},
\end{equation}
is called the Stiefel manifold.
Here $ M(n, m; \mathbb{C}) $ 
is the set of $ n \times m $ complex matrices
and $ I_k $ is the $ k \times k $ unit matrix. % of order $k$. 
The $ k $-frame
$ V = ( |v_1 (\lambda) \rangle, \ldots, |v_k (\lambda) \rangle ) $
spans a $ k $-dimensional subspace in $ \mathbb{C}^N $. The set of
$k$-dimensional subspaces is the Grassmann manifold
\begin{equation}
	G_{N, k}(\mathbb{C})
	= \{ P \in  M(N,N;\mathbb{C}) \, | \, 
	P^2=P, \, P^{\dagger} = P, \, \mathrm{tr}P=k \}.
\end{equation}
The Grassmann manifold is regarded
as the control manifold in the context of HQC.
A projection map
$ \pi : S_{N, k} (\mathbb{C}) \to
	G_{N, k} (\mathbb{C}) $
is defined as
\begin{equation}
	\pi:V \mapsto \pi(V) = VV^{\dagger}.
\end{equation}
The group $U(k)$ acts on $S_{N, k}(\mathbb{C})$ from the right 
via matrix product as
\begin{equation}
	S_{N, k}(\mathbb{C}) \times U(k) \to S_{N, k}(\mathbb{C}), \;
	(V, h) \mapsto Vh.
\end{equation}
Note that this action satisfies $\pi(Vh) = \pi(V)$.
Thus the set
$ ( S_{N, k}(\mathbb{C}), G_{N, k}(\mathbb{C}), $ $ \pi, U(k) ) $
forms a principal fiber bundle with the structure group $ U(k) $.
The group $ U(N) $ also acts on the manifolds from the left as
\begin{eqnarray}
	&&
	U(N) \times S_{N, k}(\mathbb{C}) \to S_{N, k}(\mathbb{C}), \;
	(g,V) \mapsto gV,
	\\
	&&
	U(N) \times G_{N, k}(\mathbb{C}) \to G_{N, k}(\mathbb{C}), \;
	(g,P) \mapsto gPg^{\dagger}.
\end{eqnarray}
This action is an automorphism of the principal bundle satisfying
$ \pi(gV) = g \pi(V) g^{\dagger} $.
In the Stiefel manifold the connection one-form
$ A = V^{\dagger} \, dV $ with the matrix elements
\begin{equation}
	A_{ij} (\lambda) 
	= 
	\sum_\mu 
	\left\langle v_i (\lambda)
	\left| \frac{\partial}{\partial \lambda^\mu} \right|
	v_j (\lambda) \right\rangle \, d \lambda^\mu
	\qquad
	(i,j=1,\ldots,k),
\end{equation}
is defined.
% which takes values in the Lie algebra $ \mathfrak{u} (k) $.
Then the Wilczek-Zee holonomy \cite{wz} associated with the curve $ V(t) $ 
is given by
$ %\begin{equation}
	U = \mathcal{P} e^{ - \int A },
$ %\end{equation}
where $\mathcal{P}$ denotes the path-ordered product.
% and the curvature two-form
% \begin{equation}
%	F = dA + A\wedge A= dV^{\dagger} \wedge (I_N-VV^{\dagger} )dV.
% \end{equation}

%\section{Isoholonomic problem in HQC}

We now state the main problem:
Given a unitary matrix $ U \in U(k) $ 
find the shortest closed loop $ P(t) = \pi(V(t)) $ in the Grassmann manifold
that yields the matrix $ U$ as its associated holonomy.
This problem is called the isoholonomic problem
and various representations 
of the problem have been given by Montgomery \cite{mont}.
Let us consider a familiar example,
a two-dimensional sphere,
to illustrate our idea to find the solution.
Suppose we parallel transport
a tangent vector on the unit sphere along a loop starting from and ending
at the North Pole, where the parallel transport is defined with respect
to the Levi-Civita connection \cite{gtp}. 
In general, the vector is get rotated
from the initial direction when it comes back to the initial point.
What is the shortest loop which implements the given rotation 
angle $ \omega $? 
The holonomy angle in this case is equal to the area
surrounded by the loop
and the problem is reduced to the so-called isoperimetric problem.
The solution is well-known to be a small circle which surrounds an area of $ \omega $.
We use this analogy to find a small circle solution to the isoholonomic problem in HQC.

Here we describe a generalized small circle in the Grassmann manifold.
Take 
\begin{equation}
	V_0 = 
	\left(\begin{array}{c}
		I_k\\ 0
	\end{array} \right)
	\in S_{N, k}(\mathbb{C})
	\label{eq:v0}
\end{equation}
as a reference point
and put 
$ P_0 =\pi(V_0) = V_0 V_0^{\dagger} \in G_{N, k}(\mathbb{C}) $
as the initial point of a curve in the control manifold.
% We may take without loss of generality. % correct
Taking an antihermitian matrix $ X \in \mathfrak{u}(N) $,
we define a curve $ V(t) = e^{tX} V_0 $ $ (0 \le t \le 1 ) $
in $ S_{N,k}(\mathbb{C}) $
and define 
a curve $ P(t) = \pi(V(t)) = e^{tX} P_0 e^{-tX} $ in $G_{N,k}(\mathbb{C})$ 
by projecting $ V(t) $ into $ G_{N, k}(\mathbb{C}) $.
We call the curve $ P(t) $ a small circle
if it satisfies $ P(1) = P(0) $. 
The connection evaluated along the curve $ V(t) $ is
% in $S_{N, k}(\mathbb{C})$ is
$
	V_t^*( A ) = V(t)^{\dagger} dV(t) = V_0^{\dagger} X V_0 \, dt,
$
from which we obtain the holonomy associated with the loop $P(t)$ as
\begin{equation}
	U = e^{-\int V^* A} = e^{-V_0^{\dagger} X V_0} \in U(k).
	\label{eq:holo}
\end{equation}
Note that the integrand in the exponent has no $t$-dependence and
$\int_0^1 dt$ simply yields a factor unity.

Suppose that
we are to implement a unitary gate $ U_{\rm gate} $ by HQC. 
Our task is to find a control matrix $ X \in \mathfrak{u}(N) $ 
such that the holonomy (\ref{eq:holo}) 
reproduces $ U_{\rm gate}$.
In general, computing the holonomy $ U $ for a given $ X $ is easy 
but the inverse problem is considerably difficult;
we need to find $ X $ for a given $ U $ 
while keeping the closed loop condition $ P(1)=P(0) $ satisfied.
Here we report some of the
exact solutions of this inverse problem for several important gates.

Define the logarithmic matrix $ \Omega \in \mathfrak{u}(k) $
of the gate such that $ U_{\rm gate} = e^{-\Omega} $.
If the eigenvalues of 
$ U_{\rm gate} $ are 
$ (e^{-i \omega_1}, e^{-i \omega_2}, \ldots, e^{-i\omega_k})$, 
$ \Omega $ has eigenvalues 
$ (i\omega_1, i \omega_2, \ldots, $ $ i \omega_k ) $
in the range $ -\pi < \omega_j \leq \pi$
and the same eigenvectors as $ U_{\rm gate} $. 
Then the problem (\ref{eq:holo}) reduces to find 
$ X \in \mathfrak{u}(N) $ such that
\begin{equation}
	V_0^{\dagger} X V_0 = \Omega.
\end{equation}
The general solution to this problem takes the form
\begin{equation}
	X = \left( 
	\begin{array}{cc}
		\Omega & W\\
		-W^{\dagger} & Z
	\end{array} \right)
	\label{X}
\end{equation}
with matrices
$ Z \in \mathfrak{u}(N-k) $ and $ W \in M(k, N-k; \mathbb{C})$.
Montgomery \cite{mont} has shown that
any optimal solution necessarily satisfies $ Z = 0 $.
The remaining problem is to find the matrix $ W $
that satisfies the loop condition
$ P(1) = P(0) $.
This condition demands that $ e^X $ be of the form
\begin{equation}
	e^X = \left(\begin{array}{cc}
	\ast & 0\\
	0 & \ast \ast
	\end{array} \right),
	\label{closedness}
\end{equation}
where $ \ast $ and $ \ast\ast $ are nonvanishing matrices.
The solution $ W = 0 $ is not acceptable since
it gives a constant curve $ P(t) \equiv P_0 $,
leaving the Hamiltonian unchanged.
To seek a nontrivial solution
we introduce a penalty function $ p(X) $, 
which measures the norm of the off-diagonal-block elements of 
$ e^X $, by
% mark
\begin{equation}
	p(X)  =
	\sum_{i=1}^k \, \sum_{j=k+1}^N
	\Big| \langle i | e^X | j \rangle \Big|^2,
	\label{p(X)}
\end{equation}
$ \{ |i \rangle \} $ being the set of orthonormal basis vectors of $\mathbb{C}^N$.
Then the closed loop condition (\ref{closedness}) 
is rephrased as an equivalent % mark
condition $ p(X) = 0 $. % of the penalty function.
It is apparent that $ p(X) \ge 0 $ by definition
and this nonnegativity is suitable for
numerical search of the zeroes of $ p(X) $.
Through numerical studies we have arrived at a method 
to construct systematically exact solutions of the equation $ p(X) = 0 $.

Here we describe the method briefly.
We restrict ourselves to the cases such that $ N = k + 1$ for simplicity.
In this case the Grassmann manifold $ G_{N, k}(\mathbb{C}) $
reduces to the complex projective space 
$ \mathbb{C}P^{N-1} = G_{N, N-1}(\mathbb{C}) $.
A given unitary gate $ U_{\rm gate} = e^{-\Omega} $
has a set of eigenvectors and eigenvalues 
$ \{ (u_j , e^{-i \omega_j} ) \, | \, u_j \in \mathbb{C}^k, $
$ \omega_j \in \mathbb{R}, $
$ - \pi < \omega_j \le \pi, $
$ \Omega u_j = i \omega_j u_j, $
$ \langle u_j, u_l \rangle = \delta_{jl} \} $.
Then choose a pair of eigenvector and eigenvalue, 
$ (u_\mu, \omega_\mu ) $, and substitute
\begin{equation}
	W_{(\mu,n)} = a_{(\mu,n)} u_\mu,
	\qquad
	a_{(\mu,n)} 
	= \frac{1}{2} \sqrt{(2 \pi n + \omega_\mu)(2 \pi n - \omega_\mu) }
	\label{construction}
\end{equation}
into $ W $ of $ X $ in (\ref{X}) for
$ n = \pm 1, \pm 2, \dots $.
Then we obtain the solution $ X_{(\mu,n)} $ that satisfies 
$ p( X_{(\mu,n)} ) = 0 $.
Therefore,
the integer $ n $ generates a family of solutions
and 
there are inequivalent families of solutions
as many as different eigenvalues of $ U_{\rm gate} $.
The proof of the above method is lengthy and will be published elsewhere.
In the rest of this Letter
we show results exemplifying this method.
% mark

%\section{Examples}

%\subsection{Hadamrd gate}

% Let us work out a few examples. 
We first consider the Hadamard gate
\begin{equation}
	U_{\rm H} = 
	\frac{1}{\sqrt{2}}
	\left( \begin{array}{cc}
		1&1\\
		1&-1
	\end{array} \right),
%	= e^{-\Omega_{\rm H}}.
\end{equation}
which has two eigenvalues and corresponding eigenvectors
%\begin{eqnarray}
\begin{equation}
%	&&
	e^{-i \omega_1} = 1, 
	\ u_1 = 
	\left( \begin{array}{c}
		\cos \frac{\pi}{8} \\
		\sin \frac{\pi}{8}
	\end{array} \right),
%	\\
%	&&
\quad e^{-i \omega_2} = -1, 
	\ u_2 = 
	\left( \begin{array}{c}
		-\sin \frac{\pi}{8} \\
		\cos \frac{\pi}{8}
	\end{array} \right).
\end{equation}
%\end{eqnarray}
% (frontier of editting $ \blacksquare $)
Hence there are two families of exact solutions of the equation $ p(X) = 0 $.
%and hence there are two families of solutions.
% which is defined by (\ref{p(X)}) with (\ref{X}).
% mark
We construct the first family by 
substituting the eigenvector and eigenvalue $ (u_1, \omega_1 ) $ 
into (\ref{construction}).
Then the control matrix $ X$ of the first family is
\begin{equation}
	X_{\rm H}^{(1)} 
%	= \left( \begin{array}{cc}
%		\Omega_[rm H} & W \\
%		-W^{\dagger} & 0
%	\end{array} \right)
	=
	i \pi 
	\left( \begin{array}{ccc}
		\sin^2\frac{\pi}{8} & 
		- \sin \frac{\pi}{8} \cos \frac{\pi}{8} \: &
		n e^{i \theta} \cos \frac{\pi}{8}
		\\
		- \sin \frac{\pi}{8} \cos \frac{\pi}{8} &
		\cos^2\frac{\pi}{8} &
		n e^{i \theta} \sin \frac{\pi}{8}
		\\
		n e^{-i \theta} \cos\frac{\pi}{8} & 
		n e^{-i \theta} \sin\frac{\pi}{8} & 
		0
	\end{array} \right),
\end{equation}
where $ \theta $ is an arbitrary real number,
which parametrizes unitarily equivalent solutions.
The integer $ n $ counts the winding number of the loop 
$ P(t) = e^{tX} P_0 e^{-tX} $ as seen below.
The eigenvalues and eigenvectors of $ X_{\rm H}^{(1)} $ 
are easily found to be
\begin{eqnarray}
	x_1 
		&=& i \pi, \quad 
		| x_1 \rangle = 
		\left( \begin{array}{c}
		-\sin \frac{\pi}{8} \\ 
		\cos \frac{\pi}{8} \\ 
		0
		\end{array} \right) 
	\nonumber\\
	x_2 
		&=& i n \pi, \quad
		| x_2 \rangle = \frac{1}{\sqrt{2}}
		\left( \begin{array}{c}
		e^{i \theta} \cos \frac{\pi}{8} \\ 
		e^{i \theta} \sin \frac{\pi}{8} \\ 
		1
		\end{array} \right)
	\\
	x_3 
		&=& -i n \pi, \quad
		| x_3 \rangle = \frac{1}{\sqrt{2}}
		\left( \begin{array}{c}
		e^{i \theta} \cos \frac{\pi}{8} \\
		e^{i \theta} \sin \frac{\pi}{8} \\
		-1
		\end{array} \right).
	\nonumber
\end{eqnarray}
%We chose the coefficients in the matrix $ W $ to make
%the difference $ x_2 - x_3 $ %= 2 \pi i n $
%an integral multiple of $ 2 \pi i $
%and to make the curve $ P(t) $ a closed as a result.
% mark
Using the spectral decomposition of $  X_{\rm H}^{(1)} $ 
we can calculate its exponentiation $ e^{tX} $, from which
we obtain the penalty function as
\begin{equation}
	p( t X_{\rm H}^{(1)} ) 
%	\equiv \|V_0^{\dagger} e^{tX} V_{\perp} \|^2 
	= \sin^2 (n \pi t).
	\label{eq:pen1}
\end{equation}
This clearly shows that 
the loop $ P(t) $ passes through the initial point $P_0$ $|n|-1$ times
in the interval $ 0 < t < 1 $.
% before it hits $P_0$ at $t=1$. 
Accordingly, the number $n$ describes how many times % correct
the loop winds as $t$ changes from 0 to 1. It is clear that 
$ n = \pm 1$ 
is nontrivial and optimal in this family of solutions. 

The second family of solutions is constructed by
substituting the second eigenvector and eigenvalue 
$ (u_2, \omega_2) $ of $ U_{\rm H} $
into (\ref{construction}) as
% mark
% The second family of solutions takes the form
\begin{equation}
	X_{\rm H}^{(2)}
%	= \left( \begin{array}{cc}
%	\Omega & W \\
%	-W^{\dagger} & 0
%	\end{array} \right)
	=
	i \pi
	\left( \begin{array}{ccc}
		\sin^2\frac{\pi}{8} & 
		- \sin \frac{\pi}{8} \cos \frac{\pi}{8} \: &
		- a_n e^{i \theta} \sin \frac{\pi}{8} 
		\\
		- \sin \frac{\pi}{8} \cos \frac{\pi}{8} & 
		\cos^2\frac{\pi}{8} &
		a_n e^{i \theta} \cos \frac{\pi}{8}
		\\
		-a_n e^{-i \theta} \sin \frac{\pi}{8} \: & 
		 a_n e^{-i \theta} \cos \frac{\pi}{8} & 
		0
	\end{array} \right),
\end{equation}
where $ a_n = \sqrt{4n^2-1}/2 $.
% with integer $ n $. 
The eigenvalues and corresponding eigenvectors 
of $ X_{\rm H}^{(2)} $ are
\begin{eqnarray}
	x_1 
		&=& 0, \quad
		|x_1 \rangle = \frac{1}{\sqrt{2}}
		\left( \begin{array}{c}
		\cos \frac{\pi}{8} \\
		\sin \frac{\pi}{8} \\
		0
		\end{array} \right)
	\nonumber \\
	x_2 
		&=& i \pi \left( n + \frac{1}{2} \right) \! , \quad
		| x_2 \rangle = \left( \begin{array}{c}
		-b_n e^{i \theta} \sin\frac{\pi}{8}   \\
		 b_n e^{i \theta} \cos \frac{\pi}{8} \\
		c_n 
		\end{array} \right) 
	\\
	x_3 
		&=& i \pi \left( -n +\frac{1}{2} \right) \! , \quad
		| x_3 \rangle = 
		\left( \begin{array}{c}
		-c_n e^{i \theta} \sin \frac{\pi}{8} \\
		 c_n e^{i \theta} \cos \frac{\pi}{8} \\
		- b_n 
	\end{array} \right),
	\nonumber
\end{eqnarray}
where $ b_n = \sqrt{(2n+1)/4n}$ and $c_n = \sqrt{(2n-1)/4n} $.
%We chose the value of the coefficient $ a_n $ 
%to make the difference $ x_2 - x_3 $ %= 2 \pi i n $
%an integral multiple of $ 2 \pi i $.
% mark
This time
the penalty function is evaluated as
\begin{equation}
	p( t X_{\rm H}^{(2)} ) 
	= \left(1-\frac{1}{4n^2}\right) \sin^2(n \pi t).
	\label{eq:pen2}
\end{equation}
The integer $ n $ is again interpreted as the winding number of the loop
and
the choice $ n = \pm 1 $ provides the optimal solution in this family.

We have shown that there are two families of exact solutions to
the holonomic implementation of the Hadamard gate. 
To determine the shortest loop we need to calculate length of each loop
employing the Fubini-Study metric of the Grassmann manifold.
Consequently,
the velocity of the point $ P(t) $
and therefore the length of the loop
are proportional to the norm $ \|W\| $ of the matrix $ W $.
These norms are evaluated to be
\begin{equation}
	\| W_{\rm H}^{(1)} \| = \pi |n|, 
	\qquad
	\| W_{\rm H}^{(2)} \| = \frac{\pi}{2} \sqrt{4n^2 - 1}.
	\label{W norm}
\end{equation}
Therefore, we conclude that
the simple loop solution $ n=\pm 1 $ in the second family
is in fact optimal in the whole class of solutions.

%\subsection{CNOT gate}

Next we turn to the CNOT gate
\begin{equation}
	U_{\rm CNOT} = 
	\left( \begin{array}{llll} %cccc}
	1 \: &0 \: &0 \: &0\\
	0&1&0&0\\
	0&0&0&1\\
	0&0&1&0
	\end{array} \right).
%	= e^{-\Omega_{rm CNOT}}
\end{equation}
For this case there are two families of solutions.
The first one is given by
\begin{equation}
	X_{\rm CNOT}^{(1)} 
%	= \left( \begin{array}{cc}
%	\Omega & W \\
%	-W^{\dagger} & 0
%	\end{array} \right)
	= 
	i \pi \left( \begin{array}{ccccc}
	0 & 0 & 0 & 0 & n d_1 e^{i \theta_1} \\
	0 & 0 & 0 & 0 & n d_2 e^{i \theta_2} \\
	0 & 0 & -\frac{1}{2} &\frac{1}{2} & 
	\frac{1}{\sqrt{2}} n d_3 e^{i \theta_3}\\
	0 & 0 & \frac{1}{2} & -\frac{1}{2} & 
	\frac{1}{\sqrt{2}} n d_3 e^{i \theta_3} \\
	n d_1 e^{-i \theta_1} &
	n d_2 e^{-i \theta_2} &
	\frac{1}{\sqrt{2}} n d_3 e^{-i \theta_3} &
	\frac{1}{\sqrt{2}} n d_3 e^{-i \theta_3} & 0
	\end{array} \right)
\end{equation}
with real numbers 
$ \{ d_1, d_2, d_3 \} $ with a constraint
$ d_1^2 + d_2^2 + d_3^2 = 1 $.
The parameters 
$ \{ \theta_1, \theta_2, \theta_3 \} $ are arbitrary real numbers.
The integer $ n $ again counts the winding number of the loop.
The second one is 
\begin{equation}
	X_{\rm CNOT}^{(2)} 
	= 
	i \pi \left( \begin{array}{lcccc}
	0 \quad & 0 & 0 & 0 & 0 \\
	0 & 0 & 0 & 0 & 0 \\
	0 & 0 & -\frac{1}{2} & \frac{1}{2} & 
	- \frac{1}{\sqrt{2}} a_n e^{i \theta} \\
	0 & 0 & \frac{1}{2} & -\frac{1}{2} & 
	\frac{1}{\sqrt{2}} a_n e^{i \theta} \\
	0 & 0 &
	- \frac{1}{\sqrt{2}} a_n e^{-i \theta} \: &
	  \frac{1}{\sqrt{2}} a_n e^{-i \theta} & 0
	\end{array} \right)
\end{equation}
with $ a_n = \sqrt{4n^2-1}/2 $.
By similar comparison of the norms $ \| W_{\rm CNOT} \| $
as in (\ref{W norm}),
we conclude that
the choice $ n = \pm 1$ in the second family yields the optimal loop.

%\subsection{DFT2 gate}

Our final example is the 2-qubit discrete Fourier transform (DFT2) gate
\begin{equation}
	U_{\rm DFT2} 
	= \frac{1}{2}\left( \begin{array}{cccc}
	1&1&1&1\\
	1&i&-1& -i\\
	1& -1& 1& -1\\
	1& -i& -1& i
	\end{array} \right).
%	= e^{-\Omega_{rm DFT2}}
\end{equation}
There are three families of the solutions for this gate.
The first family takes the form
\begin{equation}
	X_{\rm DFT2}^{(1)} 
	= 
	i \pi \left( \begin{array}{ccccc}
	-\frac{1}{4}& \frac{1}{4}& \frac{1}{4}& \frac{1}{4} & n w_1
	\vspace{.1cm} \\
	 \frac{1}{4}&-\frac{1}{2}&-\frac{1}{4}& 0           & n w_2
	\vspace{.1cm}\\
	 \frac{1}{4}&-\frac{1}{4}&-\frac{1}{4}&-\frac{1}{4} & n w_3
	\vspace{.1cm}\\
	 \frac{1}{4}&           0&-\frac{1}{4}&-\frac{1}{2} & n w_4
	\vspace{.1cm}\\
	n w_1^* & n w_2^* & n w_3^* & n w_4^* & 0
	\end{array} \right)
\end{equation}
with the parameters
\begin{eqnarray}
&&	w_1 =
		  \frac{1}{2}        f_1 e^{i \theta_1}
		+ \frac{1}{\sqrt{2}} f_2 e^{i \theta_2},
	\; \; \qquad
	w_2 = w_4 =
	\frac{1}{2} f_1 e^{i \theta_1},
	\nonumber \\
&&	
	w_3 =
		- \frac{1}{2}        f_1 e^{i \theta_1}
		+ \frac{1}{\sqrt{2}} f_2 e^{i \theta_2}
%	\qquad
%	w_4 = \frac{1}{2} f_1 e^{i \theta_1},
\end{eqnarray}
with a constraint $ f_1^2 + f_2^2 = 1 $.
The second family is
\begin{equation}
	X_{\rm DFT2}^{(2)} 
	= 
	i \pi
	\left( \begin{array}{ccccc}
	-\frac{1}{4} & \frac{1}{4} & \frac{1}{4} & \frac{1}{4} 
	& 0
	\\
	 \frac{1}{4} &-\frac{1}{2} &-\frac{1}{4} & 0
	& - \frac{1}{\sqrt{2}} g_n e^{i \theta}
	\\
	 \frac{1}{4} &-\frac{1}{4} &-\frac{1}{4} &-\frac{1}{4} 
	& 0
	\\
	 \frac{1}{4} & 0           &-\frac{1}{4} &-\frac{1}{2}
	& \frac{1}{\sqrt{2}} g_n e^{i \theta} 
	\\
	0 & -\frac{1}{2} g_n e^{-i \theta} & 
	0 &  \frac{1}{2} g_n e^{-i \theta} & 0
	\end{array} \right)
\end{equation}
with $ g_n = \sqrt{16 n^2-1}/4 $.
The third family is
\begin{equation}
	X_{\rm DFT2}^{(3)} 
	= 
	i \pi
	\left( \begin{array}{ccccc}
	-\frac{1}{4} & \frac{1}{4} & \frac{1}{4} & \frac{1}{4} 
	&-\frac{1}{2} a_n e^{i \theta} 
	\\
	 \frac{1}{4} &-\frac{1}{2} &-\frac{1}{4} & 0
	& \frac{1}{2} a_n e^{i \theta} 
	\\
	 \frac{1}{4} &-\frac{1}{4} &-\frac{1}{4} &-\frac{1}{4} 
	& \frac{1}{2} a_n e^{i \theta} 
	\\
	 \frac{1}{4} & 0           &-\frac{1}{4} &-\frac{1}{2}
	& \frac{1}{2} a_n e^{i \theta} 
	\\
	-\frac{1}{2} a_n e^{-i \theta} &
	 \frac{1}{2} a_n e^{-i \theta} &
	 \frac{1}{2} a_n e^{-i \theta} &
	 \frac{1}{2} a_n e^{-i \theta} & 0
	\end{array} \right)
\end{equation}
with $ a_n = \sqrt{4n^2-1}/2 $.
The norm of the matrix $ W $ with $ n= \pm 1 $ for each family
is now evaluated as
\begin{equation}
	\| W_{\rm DFT2}^{(1)} \| = \pi, 
	\quad
	\| W_{\rm DFT2}^{(2)} \| = \pi \sqrt{ \frac{15}{16} }, 
	\quad
	\| W_{\rm DFT2}^{(3)} \| = \pi \sqrt{ \frac{3}{4} }.
\end{equation}
Thus the simple loop ($n=\pm 1$) in the third family gives
the optimal control.
% The similarity between 
% the solutions of the 2-qubit discrete Fourier transform gate
% and those of the Hadamard gate originates from
% the fact that the Hadamard gate is actually the 1-qubit %discrete 
% Fourier transform gate.

%\section{Summary and Discussion}

In this Letter we considered an ideal system 
which has full isospectral parameters as control parameters.
We found exact implementation of % in the control parameter manifold
Hadamard, CNOT and DFT2 gates with a small circle in
the complex projective space
$ \mathbb{C}P^{N-1} = G_{N, N-1}(\mathbb{C}) $.
Implementation of larger-qubit gates is 
under progress and will be published elsewhere.

A realistic system has a restricted control manifold $ M $
and a control map $ f : M \to G_{N, k}(\mathbb{C})$.
For physical realization of HQC
it is required to find an optimal loop for the holonomy 
in the pullback bundle $ f^*(S_{N, k}(\mathbb{C})) $
as discussed in
\cite{pachos1,unanyan,duan,pachos2,faoro}.
The exact solutions
we have obtained here are pulled back by $f^*$ to be loops in $ M $. 
Detailed analysis of physical realization is beyond the scope of
this Letter and will be published elsewhere.

%\section*{Acknowledgements}

MN thanks 
Martti M.~Salomaa for support and warm hospitality in the Materials Physics
Laboratory at Helsinki University of Technology, Finland.
We would like to thank J.~Pachos for drawing our attention to their
recent paper \cite{pachos3}.
ST would like to thank Japan Society for the Promotion of Science (JSPS) for
partial supports of the Grant-in-Aid for Scientific Research,
No.~15540277 and MN is grateful for partial support of Grant-in-Aids for 
Scientific Research from the Ministry of Education, 
Culture, Sports, Science and Technology, Japan (Grant No.~13135215)
and from JSPS (Grant No.~14540346).

% \appendix

% \section{}
% \label{}

\end{document}